\documentstyle[amsfonts,12pt]{article}
\def\one{1\hskip-.37em 1}

\def\half{\textstyle{\frac{1}{2}}}
\def\nine{\textstyle{\frac{9}{4}}}
\def\three{\textstyle{\frac{3}{2}}}
\def\quarter{\textstyle{\frac{1}{4}}}

\def\H{{\cal H}}

\def\ep{\epsilon}

\def\l{\lambda}

\def\ra{\rightarrow}
\def\tint{{\textstyle\int}}

\def\s{\hskip.08em}
\def\d{\partial}

\def\a{\alpha}
\def\b{\begin{eqnarray*}}     
\def\e{\end{eqnarray*}}       
\def\bn{\begin{eqnarray}}     
\def\en{\end{eqnarray}}       
\def\<{\langle}
\def\>{\rangle}

\def\no{\nonumber}

\def\{{\lbrace}
\def\}{\rbrace}

\begin{document}
\title{How far apart are classical\\ and quantum 
systems?~\footnote{Based on a talk presented at the Eighth International 
Conference on Squeezed States and Uncertainty
Relations, Puebla, Mexico, June, 2003 }}
\author{John R. Klauder
\footnote{Electronic mail: klauder@phys.ufl.edu}\\
Departments of Physics and Mathematics\\
University of Florida\\
Gainesville, FL  32611}
\date{}     
\maketitle
\begin{abstract}
As is well known, classical systems approximate quantum ones -- 
but how well? We introduce a definition of a ``distance" on classical and
quantum phase spaces that offers a measure of their separation. Such a
distance scale provides a means to measure the quality of approximate
solutions to various problems. A few simple applications are discussed.
\end{abstract}

\section*{Introduction}
The purpose of this paper is to introduce a ``distance function'' which
measures the ``distance'' between classical and quantum systems. Such a
distance measure may be used, for example, to compare two different 
approximation schemes applied to a single quantum system. Thus, one can 
decide which scheme is ``better'' in the sense measured by the 
proposed distance function.

Let us start our analysis with a brief review of what constitutes a solution
set for each regime: quantum and classical. We confine our attention to
pure states. For quantum mechancis, a pure state is determined by  a
vector $|\psi\>$ in a Hilbert space $\frak H$. In fact, the physical 
content of a pure state is uniquely
specified  by  a unit vector up to an overall phase factor. Equivalently,
a pure state is defined by a one-dimensional projection operator
which we denote by $|\psi\>\<\psi|$, leaving implicit the fact that $|\psi\>$
is a unit vector, $\<\psi|\psi\>=\||\psi\>\|^2=1$. 

The equation of motion for quantum mechanics, i.e., Schr\"odinger's
equation (with $\hbar=1$),
  \bn  i(d/dt)\s|\psi(t)\>=\H\s|\psi(t)\>\;,  \label{f4}\en
where we have assumed for simplicity a time independent Hamiltonian $\H$, 
has a solution set
  \bn S_Q\equiv\{|\psi(t)\>:\s|\psi(t)\>=e^{-i\H t}\s|\psi(0)\>\;, 
\|\s |\psi(0)\>\s\|=1,\;0\le t\le T\} \en
that depends on the initial state $|\psi(0)\>$ and the particular 
self-adjoint
Hamiltonian $\H$.

For classical mechanics, the equations of motion are Hamilton's equations
  \bn {\dot q}=\d\s H/\d p\;, \hskip1cm {\dot p}=-\d\s H/\d q\;,  
\label{f6}\en
which, for a suitable classical Hamiltonian $H=H(p,q)$, have a solution set
  \bn S_C\equiv \{(p(t),q(t)): \;{\dot q}=\d\s H/\d p,\;{\dot p}
=-\d\s H/\d q,
\;(p(0),q(0)),\;0\le t\le T\}\,.  \en

The state space of quantum mechanics is a complex Hilbert space 
(modulo phase and normalization), while, for a single degree of freedom, 
the state space of classical mechanics is a real two-dimensional 
symplectic manifold, the phase space. It is
evident that given this version of quantum and classical theories, the two
systems are so different from each other that any meaningful distance
seems hopeless to define. 

As a step in the right direction, let us recall the action principles
of the two disciplines. First, we
observe that the quantum action functional may be taken to be
 \bn I_Q=\tint[\s i\<\psi(t)|(d/dt)|\psi(t)\>-\<\psi(t)|\H|\psi(t)\>]\,
dt\;,\en
and that stationary variation of this functional with respect to 
$\<\psi(t)|$
gives rise to Schr\"odinger's equation, Eq.~(\ref{f4}). Second, we 
note that the 
classical action functional is given by
  \bn  I_C=\tint[p(t)(d/dt)q(t)-H(p(t),q(t))]\,dt\;,  \en
and that stationary variations of this functional with respect to $p(t)$ and
$q(t)$ give rise to Hamilton's equations, Eq.~(\ref{f6}). It
seems that the two theories are still difficult to relate to each other.

Coherent states provide the bridge that connects the quantum and classical 
action functionals \cite{k1}. Let $|\eta\>$ be a fairly general unit 
vector in Hilbert space subject to the modest requirement that
  $  \<\eta|Q|\eta\> =\<\eta|P|\eta\>=0$,
where $Q$ and $P$  are the usual self-adjoint Heisenberg kinematical 
variables.\
Next, we introduce the two-parameter family of coherent states, defined by
  \bn  |p,q\>\equiv e^{-iqP}\s e^{ipQ}\s|\eta\> \;, \en
for all $(p,q)\in {\mathbb R}^2$. Since $[Q,P]=i\one$, it 
readily follows that
  \bn  \<p,q|Q|p,q\>=q\;, \hskip1cm \<p,q|P|p,q\>=p\;. \en
The set of coherent states spans the Hilbert space, a fact that is implicit
in the standard resolution of unity 
which holds for any unit fiducial vector $|\eta\>$ \cite{k1,k4}. 

Let us next analyze the following question: What is the consequence of 
varying the quantum action functional
over a {\it limited} set of states, say, over just the coherent states? Thus, 
let us consider
  \bn I_Q=\int[i\<p(t),q(t)|(d/dt)|p(t),q(t)\>- 
\< p(t),q(t)|\H|p(t),q(t)\>]\,dt\;,  \en
which is readily evaluated as
  \bn I_Q=\int[p(t)(d/dt)q(t)-H(p(t),q(t))]\,dt\;,  \en
where in this expression 
  \bn H(p,q)\equiv\<p,q|\H|p,q\>\;. \label{e10} \en
Evidently, the equations of motion that follow from the quantum action 
principle subject to variation only within the set of coherent states is 
entirely
equivalent to the classical equations of motion for a classical Hamiltonian
defined by (\ref{e10}). 

The connection of the quantum and classical action functionals -- and 
hence their respective equations of motion -- that arises by using the 
coherent states 
has, at last, put the two theories within a common framework. However, this
fact does not fully resolve our situation since the action functional $I$ 
does {\it not} represent a proper distance functional.

\section*{Distance Choices}
Let us recall the three fundamental properties of a ``distance function'',
$D(1,2)$, between two elements ``1'' and ``2''. These properties are:
\vskip.2cm
$\hskip.5cm$ 1) $D(1,2)\ge0\,;\;\;D(1,2)=0\;\Longleftrightarrow\; 
``1$''$=\!``2$''$\;, $\vskip.1cm
$\hskip.5cm$ 2) $D(2,1)=D(1,2)\;,$\vskip.1cm
$\hskip.5cm$ 3) $D(1,3)\le D(1,2)+D(2,3)\;. $

Frequently, one of our systems (say ``2'') will have zero distance, and 
so $D(1,2)$ will be a function of just one argument, say ``1''. In this case 
we shall simply write $D(1)$ representing the 
distance of the element ``1'' from the ``zero'' element. For the most
part we shall deal with this simpler version, while at the end we shall
consider the more general situation of $D(1,2)$. 

There are some additional conditions that we would like to impose on any 
distance function that we adopt. Expressed in a rather casual manner, the 
additional conditions we choose are:\vskip.2cm
 $\hskip.5cm$ 1) $D({\rm true\;quantum\;solution})=0\;,$\vskip.1cm
$\hskip.5cm$ 2) $D(S;0\le t\le T)+D(S;T\le t\le T+U)=D(S;0\le t\le T+U)\;,
$\vskip.1cm
$\hskip.5cm$ 3) ${\rm lim}_{\hbar\ra0}\s D({\rm classical\;solution})=0\;. $
\vskip.05cm \noindent
In item 2, $S$ denotes the solution set in question (cf., Eqs.~(2) and (4)),
apart from the time intervals involved.

Since we have found common ground for the quantum and classical 
formulations
in Hilbert space, let us next recall some standard ``distance expressions'' 
used in Hilbert space. Restricting attention to unit vectors, we consider
the distance function induced by:\vskip.2cm
$\hskip.5cm$ 1) {\it Vector norm}\vskip.1cm
$\hskip1cm$  $ D_V(|1\>,|2\>)\equiv \||1\>-|2\>\|=\sqrt{2-\<1|2\>-\<2|1\>}\;, 
$
\vskip.15cm
$\hskip.5cm$ 2) {\it Ray vector norm}\vskip.1cm
$\hskip1cm$ $D_R(|1\>,|2\>)\equiv \inf_\a \||1\>-e^{-i\a}|2\>\|=\sqrt{2}
\sqrt{1-|\<1|2\>|}\;,$\vskip.15cm
$\hskip.5cm$ 3) {\it Operator norm}\vskip.1cm
$\hskip1cm$ $D_O(|1\>,|2\>)\equiv\||1\>\<1|-|2\>\<2|\|=
\sqrt{1-|\<1|2\>|^2}\;.$  \vskip.15cm
\noindent Here the norm of a bounded operator $B$ is defined as $\|B\|=
\sup_{|\psi\>}
\|B|\psi\>\|$ over all normalized states, $\||\psi\>\|=1$.

It follows that $0\le D_O\le D_R\le D_O+D_O^2$ as well as $0\le D_R\le D_V$.
From this we learn that the norms $D_R$ and $D_O$ are
{\it equivalent} (induce identical topologies), while for infinitesimal
distances (i.e., $D_R\ll1$), the norms $D_R$ and $D_O$ are {\it equal}.
Finally, we note that $D_V$ is {\it inequivalent} to both $D_R$ and $D_O$.
Based on these several properties, as well as reasons of simplicity,  we 
choose $D_R$ to define our
desired ``distance function''.

\section*{Choice of Distance}
We construct our distance $D$ as a continuous limit of piecewise
segments. In particular, we adopt
 \bn &&D\equiv \inf_\a\s\lim_{\ep\ra0}\sum_{l=1}^N\s\||\psi_{l+1}\>-
e^{-i(\a_{l+1}-\a_l)}\s e^{-i\H\ep}\s|\psi_l\>\|  \nonumber\\
&&\hskip.4cm=\inf_\a\s\lim_{\ep\ra0}\sum_{l=1}^N\s\|i(|\psi_{l+1}\>-
|\psi_l\>)-i(e^{-i(\a_{l+1}-\a_l)-i\H\ep}-1)|\psi_l\>\s\|\;,  \en
where $\ep=T/N$.
More specifically, we define
 \bn D\equiv \inf_\a\int_0^T\|i(d/dt)|\psi(t)\>-[{\dot\a}+\H]\s
|\psi(t)\>\s\|\,dt\;. \label{e12} \en 
Here $\a(t)$, $0\le t\le T$,  represents a function over which the 
infimum takes place.
Equation (\ref{e12}) represents the basic definition introduced in this 
paper.

We immediately see that $D({\rm true\;quantum\;solution})=0$, and
$D(S;0\le t\le T)+D(S;T\le t\le T+U)=D(S;0\le t\le T+U)$.

\subsection*{Canonical examples}
Let us discuss next the ``distance'' appropriate to several 
classical systems. 
The general expression becomes
 \bn && D=\inf_\a\int_0^T\|i(d/dt)|p,q\>-[{\dot\a}+\H(P,Q)]\s|p,q\>
\s\|\,dt\no\\
&&\hskip.4cm=\inf_\a\int_0^T\|\{{\dot q}(P+p)-{\dot p}Q-{\dot\a}-
\H(P+p,Q+q)\}\s|\eta\>\s\|\,dt\;.  \en
 
As a first example, consider the harmonic oscillator with unit mass and 
unit
angular frequency, i.e., let us choose $\H=\half(P^2+Q^2)$. Thus we deal 
with
\bn &&D=\inf_\a\int_0^T\|\{{\dot q}(P+p)-{\dot p}Q-{\dot\a}-\half(p^2+q^2)
\no\\
 &&\hskip2.5cm -pP-qQ-\half(P^2+Q^2)\}\s|\eta\>\s\|\,dt\;. \en
The least value of $D$ arises if we choose 
 \bn {\dot q}=p,\hskip.4cm {\dot p}=-q,\hskip.4cm\a=\half(pq-t),\hskip.4cm
(P^2+Q^2-1)|\eta\>=0\;, \en
which leads to $ D(\rm classical\;harmonic\;oscillator)=0$,
as may have been anticipated.

More interesting is an anharmonic oscillator such as $\H=\half P^2+
\quarter Q^4$. Let us initially choose
 \bn {\dot q}=p,\hskip.3cm {\dot p}=-q^3,\hskip.3cm{\dot\a}=p\s{\dot q}-
c-\half p^2-\quarter q^4 +u,\hskip.3cm(\H-c)|\eta\>=0\;,  \en
where $u$ remains to be determined. This choice leads to 
 \bn &&D=\inf_u\int_0^T\|\{u+qQ^3+\three q^2Q^2\}|\eta\>\|\,dt\no\\ 
    &&\hskip.46cm=\inf_u\int_0^T[u^2+q^2\<Q^6\>+\nine q^4\<Q^4\>+
3uq^2\<Q^2\>]^{1/2}\,dt\no\\
   &&\hskip.46cm=\inf_u\int_0^T[(u+\three q^2\<Q^2\>)^2+q^2\<Q^6\>+
\nine q^4(\<Q^4\>-\<Q^2\>^2)]^{1/2}\,dt\no\\
  &&\hskip.46cm=\int_0^T[q^2\<Q^6\>+\nine(\<Q^4\>-\<Q^2\>^2)]^{1/2}\,dt\;, \en
where $\<Q^r\>\equiv\<\eta|Q^r|\eta\>$ 
vanishes for the present example
if $r$ is odd.
To determine $D$ we have chosen $|\eta\>$ to be the ground
state of $\H$, a condition that fixes $c$, in principle. Observe further
for the present example that for all even and positive 
$r$, $\lim_{\hbar\ra0}\<Q^r\>=0$. Consequently, we further 
learn that
  $\lim_{\hbar\ra0}\s D({\rm classical\;anharmonic\;oscillator})=0$.

\subsection*{Spin example}
It is noteworthy that the present definition for distance may be extended 
to
other systems such as spin systems. A general class of coherent states for 
spin
systems may be chosen as 
  \bn |\theta,\phi\>\equiv e^{-i\phi S_3}\s e^{-i\theta S_2}\s|s,m\>\;, \en
where $S_3|s,m\>=m|s,m\>$ and $\Sigma_j\s S^2_j|s,m\>=s(s+1)|s,m\>$ \cite{k2}.
Normally, one considers fiducial vectors $|s,s\>$ (or $|s,-s\>$), but let
us focus on a rather uncommon choice, namely $|s,m\>=|1,0\>$ \cite{k2}. 
In a standard
representation in which $S_3$ is diagonal, these states are given by
  \bn |\theta,\phi\>=\left( \begin{array}{c} e^{-i\phi}\s \sin(\theta)/
\sqrt{2}\\\cos(\theta)\\
e^{i\phi}\s\sin(\theta)/\sqrt{2} \end{array}\right) \;. \en
We choose for our example the Hamiltonian
  \bn  \H=\l\s S_3^2 =\l\left(\begin{array}{ccc}1\;&0\;&0\\0\;&0\;&0
\\0\;&0\;&1  \end{array}\right)\;.  \en
We are then led to the distance
  \bn&& D=\inf_\a\int_0^T\|i(d/dt)|\theta,\phi\>-[{\dot\a}+\H]
\s|\theta,\phi\>\s\|\,dt \no\\
 &&\hskip.45cm =\inf_\a\int_0^T[{\dot\theta}^2+\sin^2(\theta)\s{\dot\phi}^2
+{\dot\a}^2+2{\dot\a}\l\s\sin^2(\theta)+\l^2\s\sin^2(\theta)]^{1/2}\,dt\no\\
&&\hskip.45cm=\int_0^T[{\dot\theta}^2+\sin^2(\theta)\s{\dot\phi}^2+\l^2\s
\sin^2(\theta)\cos^2(\theta)\s]^{1/2}\,dt \;.  \en

It may be thought that this example corresponds to the distance for the 
corresponding classical spin system, but that would be incorrect. This fact 
follows 
since
$|\eta\>=|s,m\>$ and in that case
  \bn  i\<\theta,\phi|\s d\s|\theta,\phi\>=m\s\cos(\theta)\,d\phi\;;  \en
therefore this kinematical factor actually {\it vanishes} when $m=0$.
Thus there is no classical set of equations of motion in the present case,
and this example illustrates how it is possible to define and use the
distance function even when the restricted set of Hilbert states 
which is chosen fails
to correspond to a conventional classical system.

\subsection*{Extended coherent states}
In another direction, one may also consider so-called extended 
coherent states
which are defined, for a single degree of freedom, by 
expressions of the form \cite{k3} 
  \bn |p,q,r,s,\ldots\>\equiv\s\cdots\s e^{isY}\s e^{irX}\s e^{-iqP}\s 
e^{ipQ}\s|\eta\>\;,  \en
where $P$ and $Q$, as usual, form an irreducible Heisenberg pair, and
$X=X(P,Q)$, $Y=Y(P,Q)$, $\ldots$ . These states already form a conventional 
set
of coherent states for $p$ and $q$; the extra variables, $r,s,\ldots$, are
not needed but may (in compact form) also be used in forming a resolution
of unity. In any case, the distance definition in the present case becomes
 \bn D=\inf_\a\int_0^T\|i(d/dt)|p,q,r,s,\dots\>-[{\dot\a}+\H]\s|p,q,r,s
\ldots
\>\s\|\,dt\;.  \en
In general, the use of  extended coherent states leads to a smaller value 
for $D$
than if one had restricted oneself only to canonical coherent states.

\section*{Pair Distance}
As our final topic we return to the issue of the distance between two 
systems. 
Let 
  \bn W_j(\a_j,t)\equiv i(d/dt)|\psi_j(t)\>-[{\dot\a}_j+\H_j]\s|\psi_j(t)\>
\;,\hskip.5cm j=1,2\;, \en
denote the essential combination frequently used above. Consider the 
expression
  \bn \|W_1(\a_1,t)-e^{-i\gamma(t)}\s W_2(\a_2,t)\s\|\;.  \en
We will define the distance between system ``1'' and system ``2'' as
  \bn D(1,2)\equiv D_{12}\equiv\inf_{\a_1,\a_2,\gamma}\int_0^T\|W_1(\a_1,t)-
e^{-i\gamma(t)}\s W_2(\a_2,t)\s\|\,dt\;. \label{e15} \en
We can bound this distance by first noting that
  \bn &&\|W_1(\a_1,t)\|+\|W_2(\a_2,t)\|\ge\|W_1(\a_1,t)-
e^{-i\gamma(t)}\s W_2(\a_2,t)\s\|\no\\
  &&\hskip1cm \ge \bigg|\|W_1(\a_1,t)\|-\|W_2(\a_2,t)\|\bigg|\no\\
  &&\hskip1cm \ge \left\{\begin{array}{ll} \|W_1(\a_1,t)\|-\|W_2(\a_2,t)\| \\
\|W_2(\a_2,t)\|-\|W_1(\a_1,t)\| \end{array} \right. \;. \en
As stated, this equation holds for all $t$ and for any choice of $\a_1$
and $\a_2$. 
Therefore, by integrating over $t$ and taking $\inf_{\a_1,\a_2,\gamma}$,
we learn that
  \bn  D_1+D_2\ge D_{12}\ge |D_1-D_2|\;.  \en
Consequently, if (say) $D_2=0$, as would be the case if $|\psi_2(t)\>=
\exp(-it\H_2)$ $|\psi_2(0)\>$, then
  \bn D_{12}=D_1\;.  \en
Thus the distance of system ``1'' to any true quantum system is just 
what we have 
called $D\, (=D_1)$ previously. On the other hand, if one enquires about the 
distance between two general classical systems, then it would be necessary 
to use 
Eq.~(\ref{e15}). Of course, for two classical systems, there
are no doubt a number of plausible distance functions that would suggest
themselves as well, and it may be useful to use the specific context to
help decide which expression to choose.

\section*{Conclusion}
In this article we have introduced a ``distance function'', Eq.~(\ref{e12}),
that measures the distance of any Hilbert space temporal path from a true
quantum solution. In particular, such a function can be used to measure the
distance of a classical solution from the corresponding quantum solution.
In Eq.~(\ref{e15}) we have introduced a compatible definition for the
distance between any two Hilbert space temporal paths. 

In this article we have confined our attention to pure states. An interesting
extension of the present work would be to study 
mixed states and their dynamical
evolution. It is conjectured that the use of density matrices and their
equation of motion in the sense of von Neumann, along with the distance
function based on the operator norm, may be suitable to define an associated
distance function in that case. 

\section*{Acknowledgments}
Thanks are expressed to Lorenz Hartmann for his detailed comments.

\end{document}